\documentclass[12pt]{article}

\usepackage{amsmath,amssymb,amsthm}

\newcommand{\vtri}{\boldsymbol{\vartriangle}}

\begin{document}

\title{\bf  
Tachyons described by infinitely degenereate massless-fields \\ 
and dark matters in the universe 
}

\author{Tsunehiro Kobayashi\footnote{E-mail: 
kobayash@a.tsukuba-tech.ac.jp} \\
{\footnotesize\it Research and Support Center on the Higher Education 
for the Hearing and Visually Impaired,}\\
{\footnotesize\it Tsukuba University of Technology}\\
{\footnotesize\it Ibaraki 305-0005, Japan}}

\date{}

\maketitle

\begin{abstract}
 
A tachyon field having a negative squared-mass $-m_t^2$ 
can be 
described in terms of massless fields degenerating infinitely 
with respect to helicities. 
This picture for tachyons does not contradict causality. 
It is seen that the tachyon vector field can be quenched from the interactions 
with matter fields, 
and the effects can be represented by a phase factor. 
The accelerated expansion of the universe 
and the dark energies are interpreted in terms of the phase factor. 

\end{abstract}

\thispagestyle{empty}

\setcounter{page}{0}

\pagebreak

\hfil\break
{\bf 1. Introduction}
\vskip5pt
The Klein-Gordon equation for a tachyon having negative squared-mass $-m_t^2$ 
($m_t=$a positive real number) 
is written as 
\begin{equation} 
[\partial^\mu \partial_\mu-m^2]\phi(r_\mu )=
[{\partial^2 \over \partial \tau^2} 
- \vtri(x,y,z)-m^2]\phi(\tau,{\vec r})=0, 
 \label{tachyon} 
\end{equation} 
where the metric tensors are taken as $\eta_{\tau \tau}=1$, $\eta_{ii}=-1$ for 
$i=x,y,z$, and $\eta_{\mu\nu}=0$ for $\mu\not=\nu$, and
 $\tau=ct$ ($c$=light verocity), $m=m_tc/\hbar$, and 
$
\vtri(x,y,z)=\partial^2/ \partial x^2+\partial^2 / \partial y^2+\partial^2 / \partial z^2.
$  
It is known that in relativistic transformations tachyons must have velocities 
exceeding the light velocity $c$, and then the existence of such particles 
breaks causality. 
The tachyon is, of course, not yet observed. 
Here an idea that the introduction of the tachyons 
does not contradict causality is proposed. 
In the idea the tachyon will be described by infinitely degenerate massless-fields. 
A vector field playing the role similar to a gauge field are included in the 
massless fields. 
It will be shown that the tachyon vector-field have a close connection with problems 
in astrophysics such as the dark matter and the accelerated expansion of 
the universe. 
The membrane in Lorentz space will also be constructed 
in terms of the wave packet of the tachyons.

\vskip5pt
\hfil\break
{\bf 2. Infinitely degenerate tachyon fields}
\vskip5pt
Eq.~\eqref{tachyon} can be divided into two equations with two dimensions
~\cite{k-gauge}, e.g.  
\begin{equation} 
[{\partial^2 \over \partial \tau^2} -
{\partial^2 \over \partial z^2}]\phi_z(\tau,z)=0
\label{t1}
\end{equation} 
and 
\begin{equation} 
[-\vtri(x,y) -m^2]A(x,y)=0.
\label{t2}
\end{equation} 
One for the two-dimensional $zt$ space is the relativistic equation of motions 
for a massless particle, of which solutions are written by the plane wave 
$e^{\pm i(p_0\tau  -pz)/\hbar}$ with $p_0=|p|$ and $-\infty<p<+\infty$, 
whereas the other for the two-dimensional $xy$ space is 
a two-dimensional Schr$\ddot {\rm o}$dinger equation 
in a negative constant potential 
for the solutions with the zero-energy 
eigenvalue $E=0$. 
The $+$ and $-$ symbols in the plane wave $e^{\pm i( p_0\tau -pz)/\hbar}$, 
respectively, correspond to the creation 
and the annihilation of the field in field theories. 
Note that 
the separation of 
the two directions perpendicular to 
the non-zero three-momentum ${\bf p}$ in the four-dimensional space can 
generally be carried out 
in terms of the four-momenta defined by $p^+=(|{\bf p}|,{\bf p})$ 
and $p^-=(|{\bf p}|,-{\bf p})$ such that 
$
\epsilon^{\mu \nu \lambda \sigma} p^+_\lambda p^-_\sigma/2|{\bf p}|^2,
$
where $\epsilon^{\mu \nu \lambda \sigma}$ is 
the totally anti-symmetric tensor 
defined by $\epsilon_{0123}=1$. 
It is trivial to write 
$
\vtri(x,y) 
$ 
in the covariant expressions in terms of these tensors. 
Hereafter the moving direction of the tachyon is taken 
to be in the $z$ direction. 
It has been shown that 
2-dimensional Schr$\ddot {\rm o}$dinger equations with the central potentials 
$V_a(\rho)=-a^2g_a\rho^{2(a-1)}$ ($a\not=0$ and  
$\rho=\sqrt{x^2+y^2}$, i.e., $x=\rho \cos \varphi$ and $y=\rho \sin \varphi$.)
have zero-energy eigenstates which are infinitely degenerate~
\cite{sk-jp,k-1,ks-pr,k-ps}. 
Eq.~\eqref{t2} is nothing but the equation for $a=1$. 
This fact indicates that the tachyon can be interpreted as  
fields with the infinite degeneracy. 
First, reconsider the argument for the two-dimensional zero-energy solutions 
with the infinite degeneracy briefly. 
Putting the function $A(x,y)_n^{\pm 0}=f^\pm_n (x,y)e^{\pm imx}$ 
into Eq.~\eqref{t2}, 
where $f^\pm_n (x,y)$ are polynomials of degree $n$ ($n=0,1,2,\cdots $), 
the following equation for the polynomials is obtained 
\begin{equation}
[\vtri(x,y) \pm2im{\partial \over \partial x}]
f^\pm_n (x,y)=0.
\label{2}
\end{equation} 
From the above equations it is seen that the relations 
$(f^-_n )^*=f^+_n $ hold for all values of $a$ and $n$. 
General forms of the polynomials 
and the introduction of the angle to the $x$ axis 
have been carried out~\cite{k-1,ks-pr}. 
It has been shown that the functions $A_n$ belong to the conjugate space of 
the nuclear space in  
the Gel'fand triplet 
${\cal S}({\cal R})\subset L^2({\cal R})\subset {\cal S}({\cal R})^\times $, 
where ${\cal S}({\cal R})$, $ L^2({\cal R})$ and ${\cal S}({\cal R})^\times $ are, 
respectively, the Schwartz space, 
a Lebesgue space and the conjugate space of ${\cal S}({\cal R})$
~\cite{sk-jp,k-1,ks-pr,k-rev,bohm,sk-nc}. 
It should be stressed that in the massless case for $m=0$ 
Eq.~\eqref{2} does not have the polynomial solutions, and also 
in the positive squared-mass case for $+m^2$ the functions $A_n$ 
have the factor $e^{\pm mx}$, instead of $e^{\pm imx}$, which 
diverges one of the limits of $x\rightarrow \pm \infty$ and then 
the solutions do not have any good nuclear space 
for the construction of Gel'fand triplet. 
That is to say, only in the negative squared-mass case 
the well-defined Gel'fand triplet can be found for 
the infinite series of the functions $A_n$. 
Since all the functions have the factors $e^{\pm imx}$, 
it can be seen that 
the zero-energy states describe stationary flows in the $xy$ plane
~\cite{sk-jp,k-1,ks-pr,k-rev}. 
In this picture the tachyon mass parameter $m$ is understood 
to be the wave number of the stationary flows in the $xy$ plane, 
and the stationary flows do not change 
under the relativistic transformations in the $zt$ space. 
In general, in the relativistic covariant expression in terms of 
the anti-symmetric tensors $\epsilon^{\mu \nu \lambda \sigma}$, 
the tachyon fields behave as massless fields 
with the four-momentum $p^+$ or $p^-$ under the four-dimensional 
relativistic transformations. 
It can be said that the tachyons are the massless fields which expand into the 
two-dimensional space perpendicular to the moving direction 
as the stationary flows with the zero-energy eigenvalue. 
The tachyon fields represent no longer particles but movements of the flows. 
Of course, they move with the light velocity as massless field, 
and then the existence of the tachyon fields does not contradict causality.

\vskip5pt
\hfil\break
{\bfseries 3. Tachyon vector-field}
\vskip5pt
The tachyon fields expressed by the above polynomials $f_n^\pm$, however, do 
not have any defintie  
properties with respect to the rotations in the $xy$ space. 
Let us study  
another representation of the infinite degeneracy 
of the zero-energy states in terms of 
the two-dimensional polar coordinate~\cite{k-rev,k-spin}. 
Eq.~\eqref{t2} can be rewritten in the polar coordinate as 
\begin{equation} 
[- {1 \over \rho }{\partial \over \partial \rho }(\rho  {\partial \over \partial \rho })
+ {1 \over \rho ^2}({{\hat L}_{z} \over \hbar})^2-m^2]R(\rho ,\varphi)=0,
\label{t3}
\end{equation} 
where 
$
{\hat L}_{z}=-i\hbar {\partial \over \partial \varphi}, 
$ 
$\rho^2=x^2+y^2$, and $\varphi$ is the angle from the positive $x$ axis. 
In general $\rho^2=x^2+y^2$ should be written as a four-dimensional scalar 
in terms of the anti-symmetric tensors $\epsilon^{\mu \nu \lambda \sigma}$ 
and the four momenta $p^\pm$. 
This equation has solutions with an infinite degeneracy with respect to 
the angular momentum ${\hat L}_{z}$. 
That is to say, putting the eigenfunctions of ${\hat L}_{z}$, i.e. 
$$
R_{l}=h_l(\rho)e^{il\phi} \ \ \ {\rm for}\ \  l=0,\pm1,\pm2 \cdots
$$ 
 into Eq.~\eqref{t3}, 
the equation for $h_l$ is given by 
\begin{equation} 
[- {1 \over \rho }{d \over d \rho }(\rho  {d \over d \rho })
+{l^2 \over \rho ^2}-m^2]h_l(\rho)=0.
\label{rho}
\end{equation} 
This is the equation for the Bessel function, and then 
the solution is given by $h_l(\rho )=J_{l}(q)$, 
where $q=m\rho $. 
Here the solutions with no singularity at $q=0$ are chosen. 
We have a unique solution for each $l$. 
The solutions again have the infinite degeneracy with respect to the 
eigenvalues of $L_z$. 
That is to say, the tachyon expressed by $R_l$ represents 
a massless field with the helicity $\hbar l$. 
Note that, since the zero-energy eigenfunctions are not in a Hilbert space 
but in the conjugate space of Gel'fand triplet, 
the normalizations of the solutions cannot generally be introduced.  
It is seen that the multiplication of the dimensionless scalar-factor 
$$
Q(\tau,z;p)={p^{\mu}r_\mu \over \hbar }
$$ 
to $R_l$ does not change the property of $R_l$ as the solution of Eq.~\eqref{tachyon}, 
namely, the functions 
$$
{p^{\mu}r_\mu \over \hbar } R_l,
$$ 
can be the solutions of Eq.~\eqref{tachyon} because of the massless condition $p^2=0$.  
It is seen that 
\begin{equation}
\chi_l =f(Q) R_l
\label{chi}
\end{equation} 
are the solutions for the tachyon fields, 
where $f(Q)$ can be differentiable functions of $Q$. 
In general $T_{l \mu \nu ...}=\partial_\mu \partial_\nu ... \chi_l$ can be 
the solutions of Eq.(3) such that  
\begin{equation} 
[-\vtri(x,y)-m^2]T_{l \mu \nu ...}=0. 
\label{G2}
\end{equation} 
Taking into account that $\chi_0$ is the scalar in the four-dimensional space-time, 
arbitrary tensors in the four-dimensional space-time can be constructed such that 
\begin{equation}
T_{0 \mu \nu ...}=\partial_\mu \partial_\nu ... \chi_0. 
\label{tensor}
\end{equation}   
 Thus the tensor fields 
 $$
 \phi_{\mu \nu ...} = T_{0 \mu \nu ...}e^{ip^{\mu}r_\mu/\hbar},
 $$ 
 can satisfy Eq.~\eqref{tachyon}, 
where the four-coordinate $r=(\tau,x,y,z)$ 
and the four-momentum $p=(|p|,0,0,p)$ are taken  
in the present model. 
Note that the factor $p^{\mu}r_\mu / | p|$ represents 
$x_-=\tau-z$ for $p=(|p|,0,0,p)$ and $x_+=\tau +z$ for $p=(|p|,0,0,-p)$ 
in the light-cone coordinates. 
In the present case, 
instead of $r$, $r-r_c$ can be put in Eq.\eqref{chi}, 
where $r_c$ is a constant four-vector. 
It can be said that the vector field $\phi_{\mu}$ describes a tachyon vector-field, 
and $T_{0 \mu}$ represents the strength of the tachyon vector-field. 

\vskip5pt
Let us study more about the tachyon vector-field strength. 
Hereafter we put $T_\mu\equiv T_{0 \mu}$. 
The anti-symmetric tensor representing the field strength 
can be introduced as same as the electromagnetic field strength such that 
\begin{equation}
F_{\mu\nu } =\partial_\mu T_{\nu }-\partial_\nu T_{\mu } .
\label{F}
\end{equation} 
The tensor obviously fulfills the equation 
\begin{equation}
\partial^\nu F_{\mu\nu }=0
\Rightarrow \partial^\nu\partial_\mu T_{\nu }-\partial^2 T_{\mu }=0 
\label{F-eq}
\end{equation} 
that corresponds to the equation in the case for the absence of source. 
It can be said that $T_\mu$ is a kind of vector potentials. 
Now we can write the equation of motion of 
a matter field having spin ${1 \over 2}$ with the vector potential 
 as 
\begin{equation}
{\bar \psi}(i\gamma^\mu (\partial_\mu -ig_c T_\mu) -M)\psi ,
\label{matter}
\end{equation} 
where $g_c$ and $M$ are, respectively, the coupling constant and 
the mass of the matter field. 
In this model the transformation similar to 
the gauge transformations of $\psi$ 
and $T_\mu$ can be introduced such that  
\begin{equation}
\psi \rightarrow \psi'=e^{ig_c\chi^T}\psi, \ \ \ 
T_\mu \rightarrow T_\mu'= T_\mu +\partial_\mu \chi^T,
\label{GT}
\end{equation} 
and  the tachyon $\phi_\mu$ is transformed as 
\begin{equation}
\phi_\mu \rightarrow 
\phi_\mu'=(T_\mu+\partial_\mu \chi^T)e^{ip^\mu r_\mu /\hbar}.
\label{tT}
\end{equation} 
Under these transformation the interaction does not change the form such that  
\begin{equation}
{\bar \psi'}(i\gamma^\mu (\partial_\mu -ig_c T_\mu') -M)\psi' .
\label{matter'}
\end{equation} 
If $\chi^T=-\chi_0$ is taken, 
$T_\mu'$ obviously vanishes, namely, $T_\mu'=0$. 
In this case the matter field seems to be free from the vector potential 
induced by the tachyon vector-field. 
Actually the tachyon vector-field is quenched such that $\phi_\mu'=0$. 
The matter field, however, has an effect of the interaction with the tachyon 
on the phase expressed by the phase factor 
$$
{\tilde F}=e^{-ig_c\chi_0}.
$$

\vskip5pt
\hfil\break
{\bf 4. Dark matters in the universe}
\vskip5pt
Let us study the effect of the phase factor. 
In the evaluations all the matter fields have 
the same phase factor $\tilde{F}$ such that 
\begin{equation}
{\tilde \psi}={\tilde F} \psi.
\label{tilde-phi}
\end{equation} 
The quantity corresponding to the four-momentum of the matter field 
$$
{\tilde p}_\mu \equiv i{1 \over 2}\hbar [{\tilde \psi}^*\partial_\mu {\tilde \psi}
  -\partial_\mu {\tilde \psi}^* {\tilde \psi}]/|{\tilde \psi}|^2.
$$ 
is evaluated as 
\begin{equation}
{\tilde p}_\mu=i{1 \over 2}\hbar [\psi^*\partial_\mu \psi
  -\partial_\mu \psi^*  \psi]/| \psi|^2 
  +{1 \over 2}\hbar g_c [\partial_\mu ( \chi_0+ \chi_0^{*})].
   \label{tilde-p}
\end{equation} 
Note that in the case of the free field 
described by a plane wave $\psi \propto e^{ip_{\psi\mu} r^\mu}$ 
the first term just gives the four-momentum $p_{\psi \mu}$ of the free motion. 
It is seen that ${\tilde p}_\mu$ for ${\tilde \psi}$ 
is different from that for $\psi$ 
by the second term 
$\hbar g_c [\partial_\mu ( \chi_0+ \chi_0^{*})]/2=
\hbar g_c Re[\partial_\mu \chi_0],$ 
which is brought by the phase factor $\tilde F$. 
The second term can be understood as the difference 
from the momentum of the field being free from the interaction with 
the tachyon vector-field. 
Here let us study a simple case for $f(Q)=Q$. 
It is important that the factor $Q=p^\mu r_\mu / \hbar$ in $\chi_0$ brings  
the term that is proportional to 
time 
$
g_c|p|ct\partial_i R_0
$ for $i=x,y$. 
Since $R_0$ has no $\varphi$-dependence, 
the derivatives give non-zero effect only for the direction of $\rho$, 
that is evaluated as 
\begin{equation} 
g_c|p|ct\partial_\rho J_0(q)=-g_c|p|ctmJ_1(q).
 \label{effect}
\end{equation} 
This means that the field ${\tilde \psi}$ moves under a force without 
time-dependence, that is to say, the motion of the field 
is the motion accelerated by the time-independent force. 
In order to recover the uniformity of the space-time it is natural to consider 
that the distribution of the tachyons are uniform and have no specific direction 
for the momentum. 
Then the average with respect to the distribution of the tachyon momentum $p_\mu $ 
eliminates the term $p_z z$ 
depending on the direction of the tachyon momentum 
included in the second term, and the remaining terms are give as 
\begin{equation} 
 -g_c|{\bar p}|ctm J_1(q),
 \label{effect}
\end{equation} 
where $|{\bar p}|$ stands for the average of the magnitude of 
the tachyon momentum. 
The universal acceleration for the radial direction remind us the accelerated 
expansion of the universe~\cite{ac-UN1,accel-Un}. 
For the short time after the birth of the universe many numbers 
of tachyons were produced as same as other particles. 
During the long time of the expansion they reached to the stable 
distribution, and have been swallowed up as the phase factor by the matter fields. 
From the relation $q=m\rho$ it can be seen that, if the tachyon mass parameter 
$m=m_tc/\hbar$ is small enough to fulfill the condition that $q=m\rho$ 
is less than the first 
zero-point of $J_1$ for $q\not=0$ all over the present universe, 
all the matter fields suffer the positive acceleration for $g_c<0$ or 
the negative one for $g_c>0$. 
In such a case the acceleration does not disturb 
the internal movement of each galaxy, 
because the sizes of the galaxies are 
very small in comparison with that of the universe, 
and therefore $J_1(q)$ can be taken as 
a constant in each galaxy. 
The effect of the phase factor can be an answer 
to the accelerated expansion of the present universe. 

\vskip5pt
Let us estimate the magnitude of the effect. 
For the present time scale corresponding to the age of the universe 
the effects are observed as positive effects 
all over the universe~\cite{accel-Un}. 
This means that $g_c$ must be negative for the matter fields like protons and neutrons, 
and $q=m\rho$ must be smaller than the first zero-point of $J_1$ but $q=0$ 
so as to satisfy the condition $J_1(q)>0$. 
By using the age of the universe ($\sim 1.4\times 10^{10}$ years) 
the order of the radius of the universe is estimated as 
$10^{26}(m)$. 
From the condition that the first zero-point of $J_1$ is less than 4, 
that is,  $q=m\rho< 4$ for $\rho\sim 10^{26}$, 
the order of the tachyon mass $m_t$ is estimated as 
\begin{equation}
m_t \leq 10^{-68}(kg) .
\label{mt}
\end{equation}
This value being nearly $10^{-38}m_e$ ($m_e$=electron mass) 
is very small in comparison with masses of other particles known 
at present. 
It should, however, be noted that 
$m_t$ is not the mass but the parameter of the potential 
for deriving the zero-energy solutions. 
In order to interpret the acceleration experimentally observed~\cite{ac-UN1,accel-Un} 
the order of Eq.~\eqref{effect} should be same as those of the momenta 
of the matter fields on the present time scale of the universe. 
Since the magnitude of $ctm$ for $t=$the age of the universe 
and also that of $J_1(m\rho )$ 
for $\rho <$the size of the universe are of the order O(1), 
the order of the momentum of Eq.~\eqref{effect} is estimated as 
$$
 -g_c \bar {|p|}\sim M \gamma c \beta, 
$$ 
where the Lorentz factors are given by $\gamma=1/\sqrt{1-\beta^2}$ 
and $\beta=v/c$ for the velocity of the matter $v$. 
The order of the dimensionless coupling constant $g_c$ is obtained as  
\begin{equation}
|g_c| \sim M\gamma c \beta /\bar {|p|}.
\label{gc}
\end{equation}
Considering that the average of the tachyon momentum $\bar {|p|}$ 
will be not very small 
and also not very large in comparison with those of the matter fields, 
we can say that 
the coupling $g_c$ has a reasonable magnitude. 

\vskip5pt
Tachyons must be created also in the growth of each galaxy. 
Such tachyons give another effect to each galaxy. 
It can be said that the effect on the momentum is very small, 
since $q=m\rho$ is very small for the galaxy scale 
and $J_1(q)$ can be taken to be nearly 0 for the such small values of $q$. 

\vskip5pt
This effect also increases or decreases the energy of the matter field 
as same as the three-momentum, which is evaluated in terms of 
the 0-component of ${\tilde p}_\mu $. 
The effect obtained as  
$- g_c |\bar p|c J_0(q)$ is positive for $g_c<0$  
as same as that on the momentum. 
By using the above estimations of Eqs.~\eqref{mt} and~\eqref{gc}, 
the order of the energy correction is seen to be the same order of that 
for the matter. 
The effect of the tachyons produced in the growth of each galaxy, however, 
cannot be neglected in the energy corrections, because 
$J_0(q)$ has the largest value at $q=0$.  
This means that the dark energies can have two origins, that is to say, 
one arises from the birth of the universe and the other does from the 
birth of each galaxy. 
The former spreads all over the universe, while the latter is trapped in the 
galaxy. 
It will be not easy to observe such tachyon fields quenched by the phase factor 
in any direct processes. 
It will be observed through gravitational effects. 
It can be said that the tachyons can be the candidate for the dark matters 
in the universe and also in galaxies.

\vskip5pt
\hfil\break
{\bf 5. Some considerable effects of tachyons in astrophysics}
\vskip5pt
The zero-energy solutions contain the $l=\pm 2$ states written  
by $R_{\pm 2}$. 
This means that massless tensor fields can be described by the zero-energy solutions. 
Consider the tensor fields
\begin{equation}
h_{\mu\nu}=g\partial_\mu \partial_\nu \chi _0,
\label{graviton}
\end{equation}
where 
$g$ is a small constant. 
It can be shown that the tensor field satisfies the equation 
\begin{equation}
\partial^2 h_{\mu\nu} - \partial^\sigma (\partial _\mu h_{\nu\sigma } 
  + \partial _\nu h_{\mu\sigma }) + \eta^{\mu\nu} h_{\mu\nu}=0
\label{grav-eq}
\end{equation}
that is the linearized equation for the small fluctuation of
gravitational field $g_{\mu\nu}$ around the Minkowski metric 
$\eta_{\mu\nu}$ such that 
$$
g_{\mu\nu}=\eta_{\mu\nu} + h_{\mu\nu}.
$$  
If $g$ is taken small enough to consider that $h_{\mu\nu}$ are 
infinitesimal quantities, the corresponding infinitesimal transformations 
of coordinates are given by 
\begin{equation}
r_{\mu}'=r_\mu + g T_\mu . 
\label{coor}
\end{equation}
The gravitational interaction can be induced through  the interactions 
with the tachyons. 
Note, however, 
that all of $h_{\mu\nu}$ will possibly be quenched by the chice of the 
phase factor eliminating the tachyon vector-field, 
if the tensor field transforms as  
$$
h'_{\mu\nu}=h_{\mu\nu}+g\partial_\mu \partial_\nu \chi^T.
$$  

\vskip5pt
The tachyon  scalar-field is not quenched in this model. 
The contribution of the tachyon scalar-field to the dark energies should be considered. 
It must, however, be stressed that the tachyon fields are not observed as 
particles. 
They will be observed as something described by 
stationary flows~\cite{sk-jp,k-1,ks-pr,k-rev}. 
If the scalar field interacts with matters is described by the vector form such that 
$ {\bar \psi}i\gamma^\mu \partial_\mu \chi_0 \psi$, 
it will possibly be quenched in the choice of the phase factor.

\vskip5pt
Note also that the coupling 
constant $g_c$ for antiparticles changes the sign from that for particles. 
This means that
the effects by the phase factor 
 are opposite 
between particles and antiparticles. 
Thus these effects 
cause the difference between the distributions 
of particles and antiparticles in the universe and also in galaxies. 
It can be said that the tachyons presented here bring some interesting standpoints 
for the important problems in astrophysics.

\vskip5pt
\hfil\break
{\bf 6. Remarks on membranes}
\vskip5pt
Finally it is mentioned that the argument presented here can be extended to the case 
for the two-dimensional potentials $V_a(\rho)$ in terms of the conformal transformations
 [1-6]. 
 Actually the cases for $a=1/2$ ($V_a(\rho)\propto \rho^{-1}$;
 Coulomb type potential) and $a=3/2$ ($V_a(\rho)\propto \rho $;
 confinement type potential) were discussed in the previous works [1,6]. 
 It can also be said that 
 the case of $a=2$ representing the parabolic potential ($V_a(\rho)\propto \rho^2$) 
 will have a certain relation with string theory. 
 Actually the two-dimensional membrane can be constructed as follows: 
 Consider the wave packet in terms of the tachyon 
such that 
\begin{equation}
\Phi_l (x,y,z,t)=\int_{-\infty}^\infty d(p/\hbar) 
f(p-p_z) \chi_{l} (x,y) e^{\pm ip(z- \tau)/\hbar}.
\label{packet}
\end{equation}
Note that in this integration both states having the energy factors 
$e^{-i|p|\tau/\hbar }$ and $e^{i|p|\tau/\hbar }$ are included. 
If  
$
f(p-p_z)=e^{-\alpha^{2}(p-p_z)^2/\hbar^2}/\sqrt{\pi \alpha}
$ 
is taken, 
the result is given as 
\begin{equation}
\Phi_l (x,y,z,t)=\chi_l (x,y)  e^{-(z\mp  \tau)^2/4\alpha^2}e^{\pm ip_z(z- \tau)/\hbar}.
\label{wp}
\end{equation}
It is seen that this wave packet is a membrane 
spreading in the $xy$ plane of which peak in the $z$ direction is at $z=\pm \tau $. 
It can be seen that the wave packet does not change the structure at all, 
that is, 
not only the stationary structure in the $xy$ plane 
but the thickness 
for the $z$ direction as well. 
The wave packet, of course, moves with the light velocity $c$.  
The thickness goes to zero 
as $\alpha$ goes to zero. 
We see that the stable membrane moving with the light velocity can be 
described by the tachyons. 
It is noted that the membrane with $l\not=0$ has a vortex 
at the origin of the two-dimensional space. 
The vortex may possibly be interpreted as a string coupling with the membrane. 
On the other hand membranes without any vortices can be described 
in terms of the zero-energy solutions given in Eq.(4). 
 The membrane of the tachyons 
 may be related to that of string theory, and possibly observed as some 
 astrophysical walls.

\pagebreak

\end{document}